\documentclass{elsart}
\usepackage{graphicx,amssymb}
\journal{Physics Letters A}
\begin{document}
\begin{frontmatter}

\title{High-Probability Quantum State Transfer in an 
Alternating Open Spin Chain  with an XY Hamiltonian}

\author{E.I.~Kuznetsova},
\ead{kuznets@icp.ac.ru}
\author{A.I.~Zenchuk \corauthref{cor}
}
\corauth[cor]{Corresponding author.}
\ead{zenchuk@itp.ac.ru}

\address{Institute of Problems of Chemecal Physics, Russian Academy of Sciences, Chernogolovka, Moscow reg., 142432, Russia} 

\begin{abstract} 
This paper concerns the problem of non-ideal state transfer along  the alternating open chain of spins $s=1/2$ with the $XY$ Hamiltonian. It is found that the state transfer along  the chain with even number  of spins $N$ ($N=4,6,8$) may be realized with high probability. 
Privilege of  even $N$ in comparison with odd $N$ is demonstrated.
\end{abstract}

\end{frontmatter}


\section{Introduction}
Transmission of a quantum state from one place to another during specific time interval (quantum state transfer) is an important problem in development of quantum communication systems. The simplest model of such transfer is  the  quantum state transfer in an open 1/2-spin chain. In this case the problem may be formulated as follows. Let the spin chain be placed in the static   magnetic field  and   all spins are directed along the field except the first one which is directed opposite to the field initially at $t=0$ (initial condition). If at some moment of time  $t_{tr}>0$ we detect  that $k$-th spin is directed opposite to the field then we say that the quantum state  
has been transfered from the first node to the $k$-th node of the spin chain. Since the total spin projection must be conserved in this experiment, all other spins will be directed along the magnetic field at $t=t_{tr}$. Only such quantum processes will be considered in this letter.  We say that the quantum state is transfered along the spin chain if it is transfered from the first 
to the last node of this chain. 

Different aspects of the quantum state transfer    problem
were studied, for instance, in  refs.\cite{FBE,B,CDEL,FR,KF,VGIZ}. 
In \cite{CDEL} the possibility of the ideal transfer (i.e. transfer with probability equal to 1) of the initial quantum state along the  homogeneous  1/2-spin chains of two- and three- nodes was shown. In order to perform the ideal state transfer along the longer chains, authors suggest to use the inhomogeneous symmetric chains with different coupling constants among the neighboring nodes. But long chains constructed in this way have two
basic disadvantages: (a) coupling constants have particular values  for each  pair of nodes of the first half of the symmetric chain so that the increase of the length requires recalculation of all coupling constants and (b) spread of coupling constants increases with increase of the length of the chain, which is hard for the practical implementation.  

The fact that inhomogeneous spin chains may resolve the problem of the ideal state transfer stimulates the deep study of such chains.  Thus, the spin dynamics in alternating spin chains (i.e. chains with two different alternating coupling constants between nearest neighbor nodes) with the XY Hamiltonian at high temperatures was studied in \cite{FR} (odd number of nodes $N$) and in \cite{KF} (even number of nodes $N$).  It was demonstrated \cite{KF} that the ideal quantum state transfer   along the alternating chain with $N=4$  may be performed  for set of different pairs of coupling constants. However, $N=4$ seams to be the maximal length of the alternating chain along which the state may be perfectly transfered, which agrees with \cite{CDEL}. The long-distance entanglement in alternating 1/2-spin chain as well as in homogeneous 1/2-spin chain with small end bonds at zero temperature was studied in \cite{VGIZ}. They found that the maximal  entanglement between ends of the long spin chain is possible only in the limit of the  exact dimerization.

This letter concerns  mainly {\it the high-probability} (rather then ideal) state transfer along the alternating 1/2-spin chains with even number of nodes in an inhomogeneous magnetic field. 
The reasoning of  the non-ideal state transfer originates from the fact that the ideal state transfer is hardly reachable because of, at least,  two following   obstacles.
\begin{enumerate}
\item
Nearest neighbor approximation has been used in study of the ideal state transfer in \cite{CDEL,FR,KF}, which is not enough  to generate the ideal state transfer in practice where all nodes interact with each other.
\item
Different coupling constants in an inhomogeneous spin chain may not be produced with absolute accuracy. 
\end{enumerate}

The  Hamiltonian of this system   in the approximation of the nearest neighbors interaction may  be written in the form
\begin{eqnarray}
{\cal{H}}=\sum_{n=1}^N \omega_n I_n +\sum_{n=1}^{N-1} 
D_{n}(I_{n,x}I_{n+1,x} + I_{n,y}I_{n+1,y}),
\end{eqnarray}
where $I_{n,\alpha}$ is the projection operator of the $n$th total spin on the $\alpha$ axis, 
$w_n$ is  the Larmor spin frequency of the $n$-th node and $D_n$ is a spin-spin coupling constant.
We set
\begin{eqnarray}
w_n=0,\;\;D_n=\left\{\begin{array}{ll}
D_1, & n=1,3,\dots\cr
D_2, & n=2,4,\dots.
\end{array}\right.
\end{eqnarray}
Using Jourdan-Wigner transformation \cite{JW}
\begin{eqnarray}
&&
I^{-}_n=I_{n,x}-i I_{n,y} = (-2)^{n-1}\left(
\prod_{l=1}^{n-1} I_{l,z}
\right) c_n,\\\nonumber
&&
I^{+}_n=I_{n,x}+i I_{n,y} = (-2)^{n-1}\left(
\prod_{l=1}^{n-1} I_{l,z}
\right) c_n^+,\\\nonumber
&&
I_{n,z}=c_n^+c_n-1/2,
\end{eqnarray}
(where $c_n^+$ and $c_n$ are creation and annihilation operators of spin-less fermions) we transform this Hamiltonian to the following one \cite{FR,KF}:
\begin{eqnarray}
&&
{\cal{H}}=\frac{1}{2}{ c}^+ D { c} ,\;\;{c}^+=(c_1^+,\dots,c_N^+),\;\;{ c}=(c_1,\dots,c_N)^t,
\\\nonumber
&&
D=\left(\begin{array}{cccccc}
0 & D_1 &0 & \cdots & 0&0\cr
D_1& 0 & D_2 & \cdots & 0&0\cr
0&D_2& 0 & \cdots & 0&0\cr
\vdots &\vdots &\vdots &\vdots &\vdots &\vdots \cr
0&0&0&\cdots&0&D_j\cr
0&0&0&\cdots&D_j&0
\end{array}\right),\;\;j=\left\{
\begin{array}{ll}
1,& {\mbox{even}}\;\;N\cr
2,& {\mbox{odd}}\;\;N
\end{array}
\right..
\end{eqnarray}
Let 
\begin{eqnarray}
|n \rangle=|\underbrace{0\dots 0}_{n-1} 1 \underbrace{0\dots 0}_{N-n} \rangle
\end{eqnarray}
be the state where $n$-th spin is directed opposite to the external magnetic field while all other spins are directed along  the field.
It was shown  \cite{KF} that 
the probability for the system to be initially in the state $|1\rangle$ and
finally in the state $|N\rangle$ is defined by the following expression:
\begin{eqnarray}\label{Probability}
P(t)=\big|\langle N|\exp(-i {\cal{H}} t) |1\rangle\big|^2=\left|
\sum_{j=1}^N u_{Nj} u_{1j} \exp(-it\lambda_j/2)
\right|^2,
\end{eqnarray}
where 
$u_{ij}$ are components of the eigenvector $u_j$  corresponding to the eigenvalue $\lambda_j$ of the matrix $D$: $Du_j=\lambda_j u_j$, $u_j=(u_{1j}\dots u_{Nj})^T$.
In the case $N=4$,  eq.(\ref{Probability}) yields \cite{KF}:
 \begin{eqnarray}\label{P}
 P&=&\frac{1}{4}\left| \left(1+\frac{\delta}{\sqrt{\delta^2+4}}\right)\sin\left(\frac{D_1 t}{2} \sqrt{\frac{2+\delta^2 - \delta\sqrt{\delta^2+4}}{2}}\right) -\right.
 \\\nonumber
 &&
 \left.  \left(1-\frac{\delta}{\sqrt{\delta^2+4}}\right)\sin\left(\frac{D_1 t}{2} \sqrt{\frac{2+\delta^2 + \delta\sqrt{\delta^2+4}}{2}}\right)
 \right|^2,\;\;\delta=\frac{D_2}{D_1}.
 \end{eqnarray}
 Values $t=\bar t$ and $\delta=\bar \delta$ corresponding to the ideal state transfer are defined by the requirement
\begin{eqnarray}\label{ideal_40}
\sin\frac{\lambda_1 \bar t}{2}=-\sin\frac{\lambda_2 \bar t}{2} =\pm 1,
\end{eqnarray}
which yields
 \begin{eqnarray}\label{ideal_4}
 &&
 D_1\bar t=\frac{2(3+4 k)\pi}{
 \sqrt{2(2+\bar\delta^2+\bar\delta\sqrt{4+\bar\delta^2}))}},\;\; \bar\delta=\frac{2|1+2k-2n|}{\sqrt{(3+4k)(1+4n)}},
 \\\label{ideal_4b}
 &&
 D_1\bar t=\frac{2(1+4 k)\pi}{
 \sqrt{2(2+\bar\delta^2+\bar\delta\sqrt{4+\bar\delta^2}))}},\;\;
 \bar\delta=\frac{2|1-2k+2n|}{\sqrt{(3+4n)(1+4k)}},\\\nonumber
 &&
 \;\;n,k=0,1,\dots .
 \end{eqnarray}  
 The minimal time interval required for the quantum state  transfer: $D_1\bar t_{min}=5.441$ for $\bar\delta=1.155$. It  corresponds to  the eqs.(\ref{ideal_4}) with $k=n=0$.

 In the next section (Sec.\ref{Section}) we study the high-probability (instead of perfect)  state transfer along the chains with even (Sec.\ref{Section:N_even}) and odd (Sec.\ref{Section:N_odd}) values of $N$. Conclusions are given in Sec.\ref{Section:Conclusions}.

\section{High-probability state transfer}
\label{Section}
In this section we study the probability of the quantum state 
transfer along  the chains with different numbers of nodes. It seamed out that 
chains with even and odd $N$ exhibit significantly different 
properties as follows from Secs.\ref{Section:N_even} and 
\ref{Section:N_odd}. We will find out that  chains with 
even $N$ are preferable for the high-probability state transfer. 

\subsection{State transfer along the chain with even $N$.}
\label{Section:N_even}

\begin{figure*}[!htb]  
 \includegraphics[width=7cm,angle=270]{
 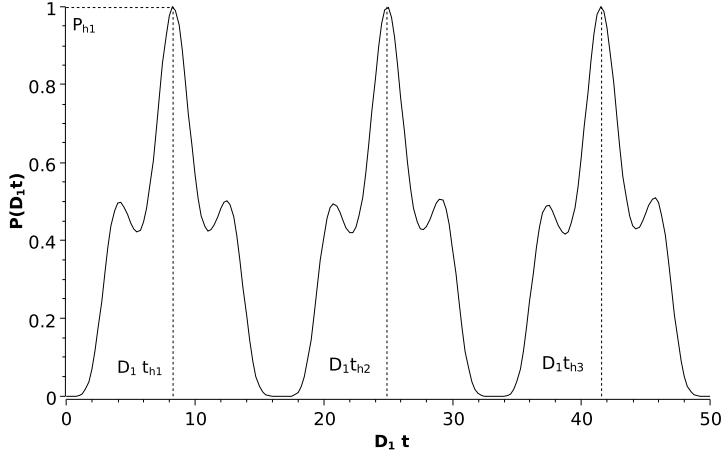}
  \caption{Probability
  of the state transfer along the four-nodes spin chain  with  ${\delta_h}_1=2.272$; ${P_h}_1=0.999$ is achieved at  $D_1 {t_h}_1=8.303$.
  }
  \label{Fig:N4_max}
\end{figure*}

\begin{figure*}[!htb]
   \includegraphics[width=7cm,angle=270]{
   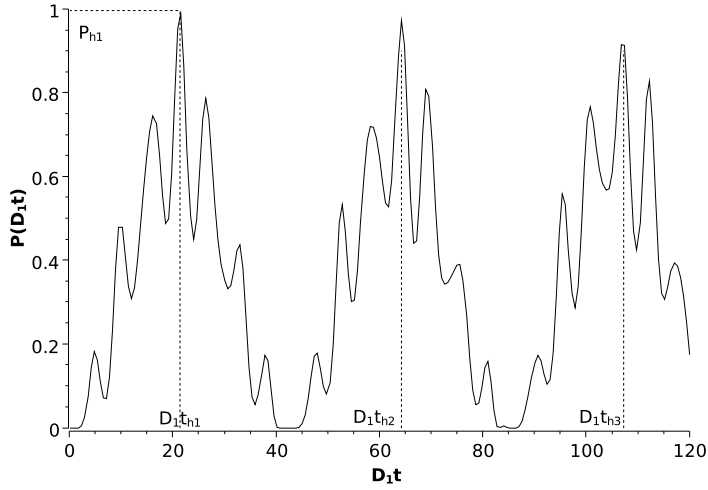}
  \caption{
 Probability of the state transfer along the six-nodes spin chain with ${\delta_h}_1=2.373$;
   ${P_h}_1=0.997$ is achieved at $D_1
{t_h}_1=21.428$.}
  \label{Fig:N6_max}
\end{figure*}

\begin{figure*}[!htb]
  \includegraphics[width=7cm,angle=270]{
  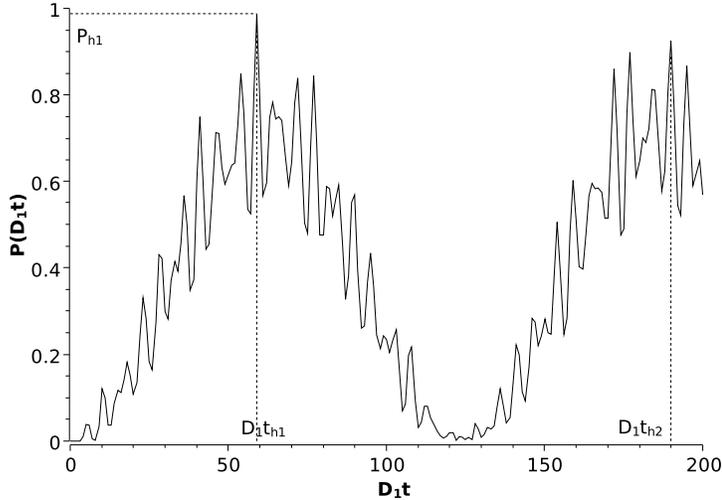}
  \caption{
 Probability of the state transfer along the eight-nodes  spin chain with   ${\delta_h}_1=2.557$; ${P_h}_1=0.989$ is achieved at  $D_1
{t_h}_1=58.966$.
  }
  \label{Fig:N8_max}
\end{figure*}

 We use some results of   \cite{KF}. Namely, consider the case $\delta=D_2/D_1 > (N+2)/N$ with $w_n=0$, $n=1,\dots,N$.
 Then  the eigenvalues $\lambda_\nu$ and components of the eigenvectors $u_{k\nu}$ with   $1\le \nu \le N$ and $\nu\neq N/2,\nu\neq N/2+1$ are given by the following expressions:
 \begin{eqnarray}\label{lambda}
\lambda_\nu &=&\left\{
 \begin{array}{ll}
 \sqrt{D_1^2 + D_2^2 +2 D_1 D_2 \cos x_\nu},& \nu=1,2,\dots,\frac{N}{2}-1,\cr
 -\sqrt{D_1^2 + D_2^2 +2 D_1 D_2 \cos x_\nu},& \nu=\frac{N}{2}+2,\dots,N
 \end{array}\right.,\\\nonumber
 u_{k\nu}&=&   \left\{
 \begin{array}{ll}
 A_\nu \sin\frac{kx_\nu}{2}, &k=2,4,\dots,N\cr
 B_\nu  \sin(N-k+1)\frac{x_\nu}{2}, &k=1,3,\dots,N-1
 \end{array}
 \right. ,\\\nonumber
 &&   
 A_\nu=\sqrt{2}\left(N+1 -\frac{\sin(N+1) x_\nu}{\sin x_\nu}\right)^{-1/2},\;\;B_\nu=A_\nu (-1)^{\nu+1},
 \end{eqnarray}
 where $x_\nu$ are solutions of the following transcendental equation, $0< x_\nu<\pi$:
 \begin{eqnarray}\label{x1}
 &&
\delta \sin\frac{N}{2} x_\nu +\sin\left(\frac{N}{2}+1\right) x_\nu=0,\;\;x_{N+1-\nu}=x_\nu,\;\;\nu=1,\dots,\frac{N}{2}-1.
 \end{eqnarray}
 If $\nu= N/2$ or $\nu= N/2+1$, then
 \begin{eqnarray}
 \label{lam_N2}
 \lambda_{N/2} &=&\sqrt{D_1^2 + D_2^2 -2 D_1 D_2 \cosh y}
,\\\nonumber
  \lambda_{N/2+1} &=&-\sqrt{D_1^2 + D_2^2 -2 D_1 D_2 \cosh y},\\\nonumber
 \label{u}
 u_{k\nu}&=&\left\{
 \begin{array}{ll}
 A_\nu (-1)^{k/2} \sinh\frac{ky}{2}, &k=2,4,\dots,N\cr
 B_\nu (-1)^{(N-k+1)/2} \sinh(N-k+1)\frac{y}{2}, &k=1,3,\dots,N-1
 \end{array}
 \right.,
 \\\nonumber
 &&
A_\nu=\sqrt{2}\left(\frac{\sinh(N+1) y}{\sinh y}-N-1\right)^{-1/2},\;\;B_\nu=A_\nu (-1)^{\nu+1},
 \end{eqnarray}
 where $y$ is a solution of the following transcendental equation, $y>0$:
 \begin{eqnarray}\label{y1}
 &&
 \delta \sinh\frac{N}{2} y -\sinh\left(\frac{N}{2}+1\right) y=0.
 \end{eqnarray}
 Due to the eqs.(\ref{lambda}-\ref{u}),  the
eq.(\ref{Probability}) may be written in the following form:
 \begin{eqnarray}\label{Probability2}
 P&=&\left|
\sum_{j=1}^{N/2} 2 u_{Nj} u_{1j} \sin(t\lambda_j/2)
\right|^2=\\\nonumber
&&2\left|\sum_{j=1}^{N/2-1} A_j^2(-1)^{j+1} 
 \sin^2\frac{Nx_j}{2}\sin(t \lambda_j/2)+\right.
 \\\nonumber
 &&
 \left.(-1)^{N/2+1} A_{N/2}^2 \sinh^2\frac{Ny}{2}\sin(t \lambda_{N/2}/2)\right|^2.
 \end{eqnarray}
In the numerical simulations below we  fix $N$ and vary $\delta$ 
with the purpose to obtain the maximum of $P$ at some moment of  time:
\begin{eqnarray}\label{P_h}
P_h=\max\limits_{\delta,t}P(\delta,t)>0.9.
\end{eqnarray}
The value $0.9$ in the RHS of (\ref{P_h}) is conventional. 
Appropriate values of $\delta$ and $t$ will be referred to as 
$\delta_h$ and $t_h$ respectively. The state transfer 
characterized by the triad $(P_h,\delta_h,t_h)$ will be referred to as
{\it high-probability state transfer}.  It is
illustrated in Figs.\ref{Fig:N4_max}-\ref{Fig:N8_max} that this triad is not unique. However, we
are interested in the high-probability state transfer having minimal
$t_h={t_h}_1$. Varying the single parameter $\delta$ we are able to  maximize 
 ${P_h}_1$. Values of other ${P_h}_i$, $i>1$, are not  important for us.

In the examples below we start with $\delta=2$ and increase $\delta$ obtaining the maximum value of ${P_h}_1$ and appropriate ${t_h}_1$. 
To anticipate, the shapes of the graphs $P(t)$ (i.e. superposition of slow and fast oscillations, see
Figs.\ref{Fig:N4_max}-\ref{Fig:N8_max}) together with 
eq.(\ref{Probability2}) suggests us to estimate ${t_h}$ in 
terms of the minimal of the eigenvalues 
$\lambda_{min}={\mbox{min}}(\lambda_1,\dots,\lambda_{N/2})$:
\begin{eqnarray}
\left|\sin \left(t_{h} \lambda_{min}/2\right)\right| \approx 1\;\;\Rightarrow\;\;
{t_h}_1\approx\frac{\pi}{\lambda_{min}}.
\end{eqnarray}
 Formulae (\ref{lambda},\ref{lam_N2}) show that $\lambda_{min}=\lambda_{N/2} $, so that
 \begin{eqnarray} 
 {t_h}_1\approx\frac{\pi}{\lambda_{N/2}}.
 \end{eqnarray}

 Thus,  for $N=4$, see Fig.\ref{Fig:N4_max}, we have found ${\delta_h}_1=2.272$, ${P_h}_1=0.999$, $D_1 t_h=8.303$.  Eigenvalues are following: $\lambda_1=2.649D_1$, $\lambda_2=0.377D_1$, so that $D_1{t_h}_1\approx\pi\frac{D_1}{\lambda_2} =8.333$. 

For $N=6$, see Fig.\ref{Fig:N6_max}, one has ${\delta_h}_1=2.373$, ${P_h}_1=0.997$, $D_1
{t_h}_1=21.428$. Eigenvalues are following: $\lambda_1=3.060D_1$, $\lambda_2=2.208D_1$, $\lambda_3=0.148D_1$, so that $D_1
{t_h}\approx \pi\frac{D_1 }{\lambda_3} =21.227$. 

 Similarly, for $N=8$, see Fig.\ref{Fig:N8_max}, one has  ${\delta_h}_1=2.557$, ${P_h}_1=0.989$, $D_1
{t_h}_1=58.966$. Eigenvalues are following: $\lambda_1=3.366D_1$, $\lambda_2=2.828D_1$, $\lambda_3=2.070D_1$, $\lambda_4=0.051D_1$. Thus $D_1
{t_h}_1\approx \pi\frac{D_1}{\lambda_4} =61.600$.

Finally we   remark    that, for practical detection of the quantum state transfer, one can select the time interval in the neighbourhood of the main peak ${P_h}_1$  by condition,  say,  $P>0.8$ everywhere inside of this time interval
and   consider that the quantum state has been transfered along the chain if it is detected at the last node during the selected time interval.

 \subsubsection{Spin chains with different $N$ and equal $\delta$.}
 We demonstrate  that the high-probability state transfer is possible along the alternating spin chains having different numbers of nodes $N$  and the same ratio of the coupling constants $\delta$.  
 We take $N=2 k$, $k=2,\dots,8$ and $\delta=2.380$. Results  are collected in the Table 1.
Disadvantage of this state transfer is the fast growth of ${t_h}_1$ with increase of $N$.

\vspace{0.4cm}
\begin{table}
\label{Table1}
\caption{}
\begin{tabular}{|p{1.2cm}|p{1.2cm}|p{1.2cm}|p{1.2cm}|p{1.2cm}|p{1.2cm}|p{1.2cm}|p{1.4cm}|}
\hline
$N$&4&6&8&10&12&14&16\\\hline
$D_1 {t_h}_1$&8.084&21.378&57.654&131.278&265.631&721.119&1403.554\\\hline
${P_h}_1$&0.990 &0.997 &0.957 &0.939&0.949& 0.962&0.901
\\\hline
\end{tabular}
\end{table}

\vspace{0.4cm}
\noindent

\subsubsection{State transfer during the given time interval.}
\begin{figure*}[!htb]  
 \includegraphics[width=7cm,angle=270]{
 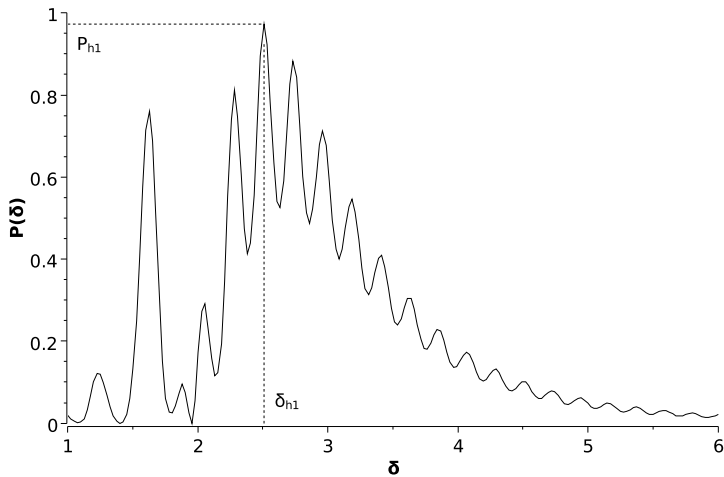}
  \caption{
  Probability of the state transfer along the
  eight-nodes spin chain at the fixed moment $D_1 {t_h}_1=60$; ${P_h}_1=0.973$ is achieved for  ${\delta_h}_1=2.510$.
  }
  \label{Fig:N8_t}
\end{figure*}

We also may arrange the high-probability  state transfer during the given time interval. For instance, let $N=8$ and suppose that  we want to transfer  the quantum state from the first node  to the last node  of the chain at $D_1 {t_h}_1=60$. Function $P(\delta)$ at $t={t_h}_1$ is represented in Fig.\ref{Fig:N8_t}. We see that it has the maximum ${P_h}_1=0.973$ at ${\delta_h}_1=2.510$. Namely the value  $\delta={\delta_h}_1$ is required for our purpose.

\subsubsection{Restrictions of the method:  the state transfer to the arbitrary  node of the chain} 
\begin{figure*}[!htb]  
 \includegraphics[width=7cm,angle=270]{
 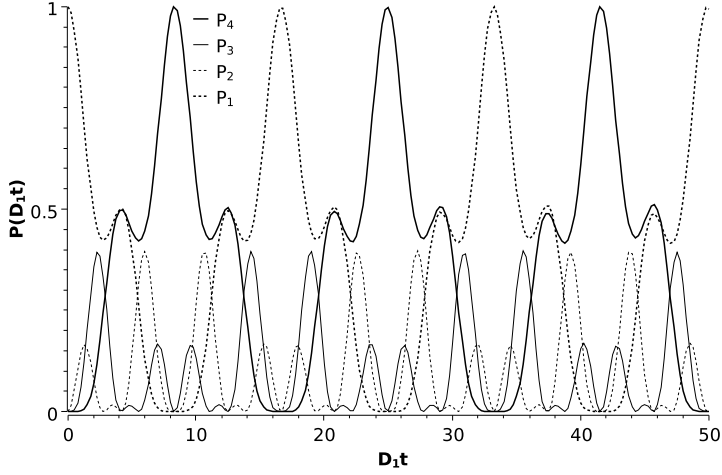}
  \caption{
 Comparison of the probabilities  $P_k$ for the state transfer to the  $k$-th node ($k=2,3,4$)  and the probability $P_1$ for the returning in the 1-st node  of the chain with  $N=4$ and $\delta=2.272$.
  }
  \label{Fig:N432_max}
\end{figure*}

The possibility to perform the high-probability state transfer between the end nodes of the spin chain suggests us to check whether the high-probability state transfer to {\it the intermediate} nodes of the chain is possible.    Expression for the probability of the state transfer to the $k$-th node of the chain, $P_k(t)$, is following:
 \begin{eqnarray}
 P_k(t)&=&\left|
\sum_{j=1}^{N} u_{kj} u_{1j}e^{-it\lambda_j/2}
\right|^2=\left\{
\begin{array}{ll}
\left|
\sum_{j=1}^{N/2} 2 u_{kj} u_{1j} \sin(t\lambda_j/2)
\right|^2,& {\mbox{even}}\;\;\; k\cr
\left|
\sum_{j=1}^{N/2} 2 u_{kj} u_{1j} \cos(t\lambda_j/2)
\right|^2,& {\mbox{odd}} \;\;\;k.
\end{array}\right. .
 \end{eqnarray}
 Unfortunately, the answer is negative at least after the numerical simulations of the state transfer along  the chains with $N=4,6,8$.
 For instance,
 graphs of $P_k$, $k=1,2,3,4$  for $N=4$ and $\delta=2.272$ (corresponding to Fig.\ref{Fig:N4_max}) are represented in Fig.\ref{Fig:N432_max}. 
 
Remark that the shapes of  the functions $P_k(t)$ illustrated in Fig.\ref{Fig:N432_max} may be interpreted as a spin-wave packet moving between the end nodes of the spin chain \cite{FBE}.

\subsection{State transfer along the chain with odd $N$.}
\label{Section:N_odd}
\label{Section:odd_N}
In this subsection we use the basic formulae derived in \cite{FR} where 1/2-spin dynamics of the chain with odd $N$ has been studied.  We set $w_n=0$, $n=1,\dots,N$. Then expressions for the eigenvalues $\lambda_\nu$  read:
\begin{eqnarray}\label{lambda_odd}
&&
\lambda_\nu=\left\{\begin{array}{ll}
D_1 \sqrt{\Delta_\nu},&\nu=1,2,\dots,\frac{N-1}{2}\cr
0,&\nu=\frac{N+1}{2}\cr
-D_1 \sqrt{\Delta_\nu},&\nu=\frac{N+3}{2},\frac{N+5}{2},\dots,N.
\end{array}\right.,\\\nonumber
&&
\Delta_\nu=1+2 \delta\cos\frac{2\pi\nu}{N+1}+\delta^2,\\\nonumber
\end{eqnarray}
Expressions  for the components of the eigenvectors $u_{j\nu}$ with $1\le \nu\le N$ and $\nu\neq (N+1)/2$ are following:  
\begin{eqnarray}\label{u_odd}
&&
u_{j\nu}=\left\{\begin{array}{ll}
\frac{A_\nu D_1}{\lambda_\nu}\left(
\delta \sin\frac{\pi \nu(j-1)}{N+1} +\sin\frac{\pi\nu(j+1)}{N+1}\right)&j=1,3,\dots,N\cr
A_\nu\sin\frac{\pi\nu j}{N+1},& j=2,4,\dots,N-1,
\end{array}\right.,\\\nonumber
&&A_\nu=
\sqrt{\frac{2}{N+1}}.
\end{eqnarray}
In addition,
\begin{eqnarray}\label{uN_odd}
\\\nonumber
&&
u_{j(N+1)/2}=\left\{
\begin{array}{ll}
B(-\delta)^{(N-j)/2},&j=1,3,\dots,N\cr
0,&j=2,4,\dots,N-1.
\end{array}
\right.,\;\;B=\sqrt{\frac{\delta^2-1}{\delta^{N+1}-1}}.
\end{eqnarray}
Eq.(\ref{Probability}) gets the next form:
\begin{eqnarray}\label{P_odd}
&&
P=\left|2\sum_{j=1}^{(N-1)/2} u_{Nj}u_{1j}\cos(\lambda_j t/2)+u_{N(N+1)/2}u_{1(N+1)/2} \right|^2=\\\nonumber
&&
\left|2\sum_{j=1}^{(N-1)/2} A_j^2\frac{D_1^2\delta}{\lambda_j^2}\sin\frac{2\pi j}{N+1}
\sin\frac{\pi j(N-1)}{N+1} \cos(\lambda_j t/2)+B^2(-\delta)^{(N-1)/2} \right|^2.
\end{eqnarray}
Behaviour of the function $P(t)$ is completely different in comparison with the case of even $N$. It was shown  \cite{CDEL} that the ideal state transfer is possible for $N=3$ and is impossible for $N>3$ if $\delta=1$. Using numerical simulations we obtain that the  high-probability state transfer is possible, in principle, only for $\delta \approx 1$.

\begin{figure*}[!htb]
   \includegraphics[width=7cm,angle=270]{
   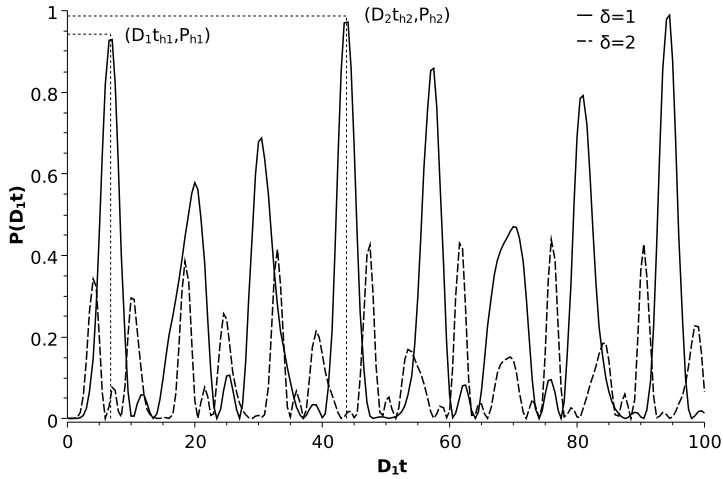}
  \caption{
 Comparison of probabilities for the state transfer along the  two five-spin chains with $\delta=1$ ($\lambda_1=1.732 D_1$, $\lambda_2=D_1$) 
  and $\delta=2$ ($\lambda_1=2.646 D_1$, $\lambda_2=1.732 D_1$) .
  }
  \label{Fig:N5_max}
\end{figure*}

For instance, $P(t)$ for $N=5$ and $\delta=1$ is represented in Fig.\ref{Fig:N5_max}. It has a set of maxima. We mark two mostly  considerable of them: ${P_h}_1=0.942$ at $D_1{t_h}_1=6.764$ and  ${P_h}_2=0.987$ at $D_1{t_h}_2=43.757$.  

One can demonstrate that the amplitude of $P$ decreases with increase of $\delta$. As an example, the functions  $P(D_1 t)$ for  $N=5$ and two values  $\delta=1$ and $2$ are represented in Fig.\ref{Fig:N5_max}.

We emphasise two following features of the high-probability  state transfer along the chain with odd $N$.
\begin{enumerate}
\item
The probability is described by an  oscillating function similar to the case of even $N$. However, unlike the case of even $N$,   the amplitude of $P$ decreases with increase of $\delta$. Because of this fact, we may not  effectively use parameter $\delta$ in order to provide the high-probability state transfer during the reasonable time interval $(0,{t_h}_1)$.
\item 
The high-probability  state transfer ($P_h\sim 0.9$) is observable in the neighbourhood of  $\delta = 1$.  However, appropriate $t_h$ may be too long for realization in quantum process.
\end{enumerate}
Thus,  we  conclude that the   chains with odd $N$ are less suitable  for  the high-probability state transfer in comparison with the  chains with even $N$.

The obtained result may be justified by the analytical estimation of $P$ for the quantum state transfer along the chain with odd $N$. Using eqs.(\ref{lambda_odd}-\ref{P_odd}) one has
\begin{eqnarray}
&&
P\le (F_1 +F_2)^2,\\\nonumber
&&
F_1=\left|2\sum_{j=1}^{(N-1)/2} A_j^2\frac{D_1^2\delta}{\lambda_j^2}\sin\frac{2\pi j}{N+1}
\sin\frac{\pi j(N-1)}{N+1} \cos(\lambda_j t/2)\right|,\\\nonumber
&&
F_2=\left|B^2(-\delta)^{(N-1)/2}\right|.
\end{eqnarray}
Consider $F_1$ and $F_2$ separately. 
For our convenience, we consider the case $\delta\ge 1$ without loss of generality. Using inequality $|\cos(\lambda_j t/2)|\le 1$ and equations (\ref{lambda_odd},\ref{u_odd})  one gets the following chain   of inequalities: 
\begin{eqnarray}\label{F1}
&&
F_1\le \frac{4}{N+1}\sum_{j=1}^{(N-1)/2} \left|
\frac{D_1^2\delta}{\lambda_j^2} \sin^2\frac{2\pi j}{N+1} 
\right|=\\\nonumber
&&
\frac{4}{N+1}\sum_{j=1}^{(N-1)/2} 
r_j(\delta)\sin^2\frac{\pi j}{N+1}
\le
\frac{2 \Delta(\delta) }{N+1} \sum_{j=1}^{(N-1)/2} 
\left(1-\cos\frac{2\pi j}{N+1}\right)
=
\\\nonumber
&&
\frac{\Delta(\delta) (N-1)}{N+1}.
\end{eqnarray}
Here
\begin{eqnarray}
&&
r_j(\delta)= \frac{\left(2+2\cos\frac{2\pi j}{N+1}\right)}
{\left(\delta+1/\delta+2 \cos\frac{2\pi j}{N+1}\right)},\;\;j=1,\dots,(N-1)/2,\;\;0\le r_j \le 1,\\\label{DELTA}
&&
\Delta(\delta)=\max\Big(r_j(\delta),\;1\le j\le (N-1)/2\Big)=r_1\\\nonumber
&&
=\frac{\left(2+2\cos\frac{2\pi}{N+1}\right)}
{\left(\delta+1/\delta+2 \cos\frac{2\pi}{N+1}\right)}\le 1.
\end{eqnarray}
Deriving (\ref{F1}) we used  the inequality $(\delta+1/\delta)\ge 2$ and the identity $\displaystyle \sum_{j=1}^{(N-1)/2} 
\cos\frac{2\pi j}{N+1}=0$.  

Next,
\begin{eqnarray}\label{F2}
&&
F_2=\delta^{(N-1)/2}\frac{\delta^2-1}{\delta^{N+1}-1}=
\frac{\delta^{(N-1)/2}}{\sum_{k=0}^{(N-1)/2}\delta^{2k}}=
\\\nonumber
&&
\left\{
\begin{array}{ll}
\left(\sum_{k=0}^{(N-3)/4}(1/\delta^{2k+1} +\delta^{2k+1})
\right)^{-1},& {\mbox{odd}}\;\;\;(N-1)/2\cr
\left(\sum_{k=1}^{(N-1)/4}(1/\delta^{2k} +\delta^{2k})
+1\right)^{-1},& {\mbox{even}}\;\;\;(N-1)/2
\end{array}
\right\}\le \frac{2}{N+1}.
\end{eqnarray}
Thus
\begin{eqnarray}\label{P_odd_l}
P\le {\cal{P}}=\left(\frac{\Delta(\delta)(N-1) +2)}{N+1}\right)^2.
\end{eqnarray}

If $\delta=1$, then $\Delta(1)=1$ so that the inequality (\ref{P_odd_l})
yields $P\le 1$, which means that the state transfer may approach
the high-probability state transfer at some moment of time $t_h$. However, $t_h$ may be too long as it was mentioned above.
 It follows from the eq.(\ref{P_odd_l}) that ${\cal{P}}$ decreases with
 increase of $\delta$.  For instance, if $\delta=2$ and $N=5$, then 
 $\Delta(2)=6/7$ and ${\cal{P}}=361/441 \approx 0.819$ which agrees with Fig.\ref{Fig:N5_max}. 
  
  Emphasize that sign $"="$ in the inequalities (\ref{F1}) and (\ref{P_odd_l}) may appear only if the following conditions are valid at some moment of time $t_0$: 
\begin{eqnarray}\label{con1}
&&
|\cos(\lambda_j t_0/2)|=1,\;\;\forall \;\;j=1,\dots, (N-1)/2,\\\label{con2}
&&
\delta=1.
\end{eqnarray}
Since the parameter $\delta$ is fixed by the condition (\ref{con2}), one has one parameter $t_0$ in order to satisfy $(N-1)/2$ conditions (\ref{con1}). This is possible if only $N=3$ when the system (\ref{con1}) reduces to the single equation \cite{CDEL}. In the case $N>3$ the sign $"\le"$ must be replaced by the sign $"<"$  in  inequalities (\ref{F1}) and (\ref{P_odd_l}).
 
  Similar to the case of even $N$, numerical simulations  show that there is no high-probability state transfer to the inner nodes of the chain.
 
 Remark that the high-probability state transfer along the chains with even $N$  may be realized  only if $N$ is not too large. In fact, $\lambda_{N/2}\to 0$  and  ${t_h}_1 \to \infty$ as 
 $N\to \infty$, while ${t_h}_1$ may not be too long in quantum 
 process. This observation removes  principal differences between long  chains with even and odd $N$ in the quantum state transfer systems.

\section{Conclusions}
\label{Section:Conclusions}
We have demonstrated that the alternating short spin chains with even numbers of nodes $N$ are preferable for the purpose of the quantum state transfer. Although the ideal state transfer for  $N>4$  is impossible in the alternating chain, the state transfer along the chain with even $N$ 
may be performed with  high probability. This interesting phenomenon is especially important because the ideal state transfer is hardly achievable in  practice.


Authors thank Prof. E.B.Fel'dman for useful discussions.
This work is supported by Russian Foundation for Basic Research through the grant 07-07-00048.

\end{document}